\documentclass[notitlepage,twocolumn,showkeys,nofootinbib]{revtex4-1}

\usepackage{amsmath}
\usepackage{amssymb}

\usepackage{graphicx}

\newcommand{\rmi}{\mathrm{i}}
\newcommand{\rmd}{\mathrm{d}}
\newcommand{\rme}{\mathrm{e}}
\newcommand{\bs}{\boldsymbol}
\newcommand{\appropto}{\mathrel{\vcenter{
  \offinterlineskip\halign{\hfil$##$\cr
    \propto\cr\noalign{\kern2pt}\sim\cr\noalign{\kern-2pt}}}}}

\allowdisplaybreaks

\begin{document}

\title{Singular knot bundle in light}

\author{Danica Sugic}

\email{sugicdanica@gmail.com}

\affiliation{H. H. Wills Physics Laboratory, University of Bristol, Bristol BS8 1TL, UK}

\author{Mark R. Dennis}

\affiliation{H. H. Wills Physics Laboratory, University of Bristol, Bristol BS8 1TL, UK}
\affiliation{School of Physics and Astronomy, University of Birmingham, Birmingham B15 2TT, UK}

\begin{abstract}
As the size of an optical vortex knot, imprinted in a coherent light beam, is decreased, nonparaxial effects alter the structure of the knotted optical singularity.
For knot structures approaching the scale of wavelength, longitudinal polarization effects become non-negligible and the electric and magnetic fields differ, leading to intertwined knotted nodal structures in the transverse and longitudinal polarization components which we call a knot bundle of polarization singularities.
We analyze their structure using polynomial beam approximations, and numerical diffraction theory.
The analysis reveals features of spin-orbit effects and polarization topology in tightly-focused geometry, and we propose an experiment to measure this phenomenon.\\

\end{abstract}

\maketitle

\section{Introduction}
\label{sec:Introduction}

Creating knotted structures \cite{Kauffman} in the complex amplitude patterns of structured light is a challenge for holographic beam shaping. 
Experiments and theory of optical knots were described in \cite{Leach2004, Dennis2010}, following earlier theory \cite{Berry2001a, Berry2001b}.
These schemes involved superpositions of laser modes---implemented holographically---containing optical vortices in configurations of various knots and links, within the focal volume of a paraxially propagating laser beam.
The study of optical vortex knots in monochromatic fields complements the study of knotted structures realized in various other physical systems, including knotted electromagnetic fields \cite{Arrayas2017}, fluid vortex lines in water \cite{Kleckner2013}, knotted small molecules \cite{Danon2017} and bio-molecules such as DNA \cite{Liu1981}, defects around colloidal structures in liquid crystals \cite{Tkalec2011,Martinez2014}, and knot-like structures in Bose-Einstein Condensates (BECs) \cite{Hall2016}.

In most of these other systems, energetic considerations affect the knots' stability, and exciting knotted structures in the physical medium is the main challenge.
Knotted vortices in light, on the other hand, have no energy associated with them; the primary technical challenge in realizing them experimentally consists in controlling the delicate parameter ranges where topological structures occur, for instance by minimizing perturbations (such as aberrations) \cite{Leach2005}.
An application of such spatial configuration of vortex knots is to template knotted structures in photosensitive materials, extending previous, simpler experiments, with imprinting structured optical amplitudes into BECs \cite{Bhowmik2016}, liquid crystals \cite{Marrucci2006} and other materials \cite{Ambrosio2012, Ni2017}.

The theory of knotted optical singularities in light beams developed in \cite{Berry2001a, Berry2001b, Leach2004, Dennis2010} was primarily in the paraxial regime; the superpositions of knots in Gaussian beams were of the order of the waist width $w$ transversally (of the order $10^3$-$10^4 \lambda$), but of the Rayleigh distance longitudinally ($\pi w^2 \lambda^{-1} \approx 10^6$-$10^8 \lambda$).
Such an extreme aspect ratio is impractical to imprint into material systems, so it is natural to adapt this approach to create smaller knotted structures in optical fields, of a scale closer to the wavelength, whose physical aspect ratio is closer to unity.

We approach this problem by exploring, analytically and numerically, the effect of reducing the transverse size of the fields constructed in \cite{Dennis2010}.
A paraxial field is one which approximately resembles a plane wave of wavelength $\lambda$ propagating in $z$.
In the paraxial regime, the electric and magnetic part of the light field can be represented by the same complex scalar amplitude function multiplying a constant, transverse polarization vector; the magnitude of the longitudinal components is negligible. 
Mathematically we show that there is, in fact, also a knotted vortex in these nearly zero components.
A small vortex knot, approaching the scale of the wavelength, is outside the paraxial regime.
The field around such a structure has a non-negligible longitudinal component and the electric and magnetic fields no longer agree with each other.

We will describe a regime of transverse knot size (of the order of several $\lambda$) where the focal energy is approximately evenly distributed between transverse and longitudinal components, and between electric and magnetic field components.
In this regime, the paraxial field structures are perturbed by nonparaxial effects, enough to deform the knots in each component, but not so strong as to destroy the nodal knot topologies.
The resulting electromagnetic field therefore displays a \emph{bundle} of intertwined, knotted polarization singularities (corresponding to the nodes in the various components) which we will describe in detail.
We also propose an experiment in which such a knotted object might be measured.
Only in the nonparaxial regime the singular bundle can be observed because, although it is already present in the paraxial regime, the electric and magnetic knots cannot be distinguished, and the longitudinal components are negligible.
The size of this nonparaxial knot bundle can range from several wavelengths---both in the transverse and longitudinal directions---to subwavelengths, which makes it ideal for light-matter imprinting.

We focus our investigation on the behavior of the simplest knotted light field, namely that around a trefoil knot with 3-fold symmetry.
As described in \cite{Dennis2010}, this knot evolves paraxially forwards and backwards from an initial condition in the focal plane $z=0$, found using a topological algorithm (a slice of a Milnor polynomial \cite{Dennis2010,Bode2017a}), multiplying a Gaussian of fixed width $w$,
\begin{equation}
\psi_{\mathrm{2D}}(r,\phi)=\mathrm{e}^{-r^2/(2w^2)}  (1 -r^2 -r^4 + r^6 -8 r^3 \mathrm{e}^{3 \rmi \phi} ).
   \label{eq:milnor}
\end{equation}
Fields with this transverse profile are an interesting class of structured beams for considering spin-orbit effects, as they are the superposition of two eigenfunctions of optical orbital angular momentum (OAM): a factor proportional to $\mathrm{e}^{\rmi n\phi}$ with azimuthal index $n$, superposed with an axisymmetric factor \cite{Romero2011}.
Assuming the polarization in the initial, focal plane is purely helical, with right or left handed circular polarization $\widehat{\bs{e}}_{\pm} = (\widehat{\bs{x}} \pm \rmi \widehat{\bs{y}})/\sqrt{2}$, the different OAM states propagate differently from each other, and according to the handedness of the transverse polarization.
The resulting polarization structure of small knots therefore depends on the interplay of spin and orbital optical angular momentum \cite{Bliokh2011, Bliokh2015}.
In particular, we will see that the choice of helicity opposite the sense of phase increase in (\ref{eq:milnor}) is important to the structure of the polarization bundle, and assume that the transverse amplitude (\ref{eq:milnor}) in the focal plane multiplies left-handed circularly polarized light $\widehat{\bs{e}}_- = (\widehat{\bs{x}} - \rmi \widehat{\bs{y}})/\sqrt{2}$.

The trefoil function (\ref{eq:milnor}) is just one of many knot functions known to give vortices in the configurations of different kinds of knots and links on propagation \cite{Dennis2010, King, Bode2017a}; the behavior of other propagating knot functions is broadly similar and we briefly consider some others here.
A convenient (but not unique) way of generating the $m,n$-torus knot/link \cite{Kauffman} (including the trefoil knot) is to take, as the initial condition, the numerator of the fraction $u^m - v^n$, where $u = (r^2 - 1)/(r^2 + 1)$, $v = 2 r \mathrm{e}^{\rmi \phi}/(r^2 + 1)$ times the Gaussian factor \cite{Dennis2010, Bode2017a, King} which are therefore superpositions of two fields with OAM indices $0$ and $n$.
The trefoil profile (\ref{eq:milnor}) has $(m,n) = (2,3)$, and $(3,2)$ gives a trefoil with a different spatial conformation.
Other simple torus knots and links include the unknot, the Hopf link $(2,2)$, the double link$(2,4)$ or $(4,2)$, the cinquefoil knot $(2,5)$ or $(5,2)$ (the Milnor polynomial of the latter requires multiplication by an extra factor $(1+r^2)$ \cite{King,Bode2017a}).

The beams we investigate are monochromatic fields, where the singular configurations are static.
When the electric field has the form $\bs{E}(r,\phi,z)\mathrm{e}^{\rmi  (k z - \omega t)}$ in cylindrical coordinates $r,\phi,z$, the vector $\bs{E} = \bs{E}(r,\phi,z)$ satisfies the reduced Helmholtz equation
\begin{equation}
   \nabla^2_{\bot} \bs{E} + \partial_z^2\bs{E} + 2 \rmi k \partial_z \bs{E} = 0,
   \label{eq:helmw}
\end{equation}
where $\nabla_{\bot}^2 \equiv \partial_r^2 + r^{-1} \partial_{r} + r^{-2} \partial_{\phi}^2$ is the transverse Laplacian.
In particular, in the paraxial regime, $k$ is large as an inverse wavelength and $|\partial_z^2\bs{E}|$ becomes much smaller than the other terms in (\ref{eq:helmw}), so $\bs{E}$ may be approximated by one satisfying the paraxial equation $\nabla^2_{\bot}\bs{E} + 2 \rmi k \partial_z \bs{E} = 0$. 
Since $\nabla\cdot(\bs{E}\mathrm{e}^{\rmi k z}) =0$, the divergence of the transverse components of $\bs{E}$ is approximately $- \rmi k E_z$, and since $k$ is large, the longitudinal component $E_z$ is negligible compared to the transverse components.
Furthermore, the magnetic field is also polarized with the same amplitude function, and the loci of the vortex lines in the $\bs{E}$ and $\bs{B}$ coincide.

Outside the paraxial approximation the longitudinal component can become comparable to the transverse components. 
In this regime, different components of $\bs{E}$ have optical vortices (nodal lines) in slightly different positions, whose singular structure is naturally described  in terms of polarization singularities: lines in 3D space with the same polarization state \cite{Dennis2009}.
The splitting of scalar vortices of transversely polarized paraxial beams into polarization singularities has been studied in some detail \cite{Berry2004c}.
The polarization singularities of the transverse component of the field---whose description does not include the longitudinal component---are C lines when the polarization is circular, and L surfaces when the polarization is linear \cite{Dennis2009}.
In the case of the full 3D-vector field, the polarization singularities are the \textit{true} points of circular and linear polarization, thus they are denoted, respectively, C\textsuperscript{T} lines and L\textsuperscript{T} lines \cite{Nye1987}.

The design of holograms for nonparaxial fields is very challenging, therefore it is extremely hard to structure arbitrary polarization singularity patterns.
The problem has been mainly addressed with numerical optimization techniques \cite{Jabbour2008} and semi-analytical approaches \cite{Braat2003, Foreman2008}, but they can be hard to implement.
In our analytical treatment, we consider the nonparaxial behavior of fields structured on the scale of wavelength, but where the Gaussian envelope width $w$ is fixed, using polynomial beam functions \cite{Dennis2011} (extending the scalar approach of \cite{Dennis2010} to the nonparaxial vector regime), including longitudinal polarization. 
For the regime in which the Gaussian envelope is made small together with the knot, the beam is propagated numerically by evaluating the Richards-Wolf vector diffraction integral \cite{RW1959, Novotny}.

The structure of this article is as follows.
In the next section, we present a physical treatment of all the electromagnetic field components of the paraxial beam and the presence of knots in all the components.
In section \ref{sec:nonparaxial} we present a theoretical analysis of the polarization knots that constitute the singular bundle, of the order of the wavelength, embedded in a beam with wide Gaussian envelope.
In section \ref{sec:experiment} we optimize the bundle regime both theoretically and numerically, leading to an experiment design where the Gaussian envelope and the transverse size of the knot are of the same order and all the components of the beam have comparable amplitudes.
We discuss and extend our results in section \ref{sec:discussion}.

\section{Paraxial polarization knots}
\begin{figure}[h]
\centering
\includegraphics[scale=1]{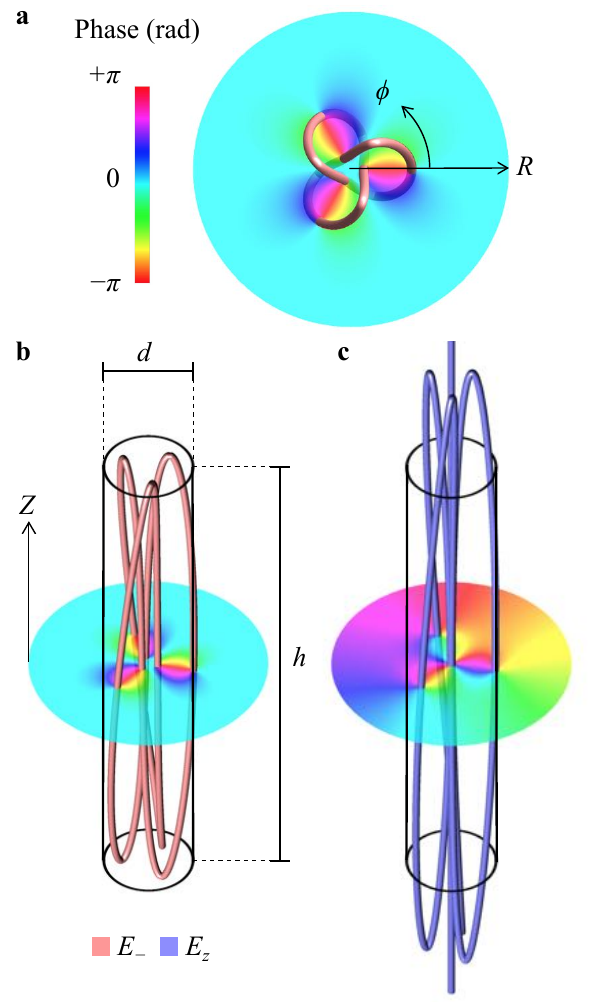}
\caption{
   \textbf{Paraxial optical vortex trefoil knots.}
   (a) Phase distribution on the plane $z=0$ of the field (\ref{eq:milnor}) which was imprinted on a hologram plate in \cite{Dennis2010} to create an optical vortex knot. 
   (b) 3D view of the trefoil vortex knot (red) in the transverse component from (\ref{eq:E-par}). 
   The cylinder of diameter $d$ and height $h$ encloses the knot structure. 
   (c) 3D view of trefoil knot and axial vortex in the longitudinal component (blue) from (\ref{eq:Ezpar}).
   The same cylinder from (b) is given for reference.
   Physically, the knots' length along the propagation direction scales quadratically with respect to their transverse size. 
}
\label{fig:Rescaling}
\end{figure}

In this section we describe paraxial beams by polynomial functions and we determine the longitudinal size of optical vortex knots with respect to their transverse length; we then include polarization features to the electric field and we track the nodal lines of each of its components in the circular basis to reveal the polarization knots.

When studying the fine structure of light, we only need to describe features of the amplitude in the region of interest.
So-called \emph{polynomial beams} \cite{Dennis2011} provide an approach to such a local study of optical amplitude structures. 
These are the formal, propagating solutions of the paraxial equation or reduced Helmholtz equation (\ref{eq:helmw}), which in the focal plane $z=0$ take the form $r^{\lvert\ell \rvert +2p}\mathrm{e}^{\rmi \ell \phi}$ in cylindrical coordinates (with integer $\ell$ and $p>0$).
Although polynomial solutions are not satisfactory global representations of a beam---not being normalized---they arise naturally from a Taylor expansion of any beam about the origin (focal point). 
Paraxial polynomial beams are analogous to the heat polynomials which solve the heat equation.
Polynomial light beams were originally introduced to study the behaviour of optical vortices by Berry \cite{Berry1998} and Nye \cite{Nye1998}, and have subsequently been studied by several groups \cite{Dennis2011,Berry2001b,Torre2011,Borghi2011}.
They can be understood as a basis for the description of \emph{superoscillatory} behavior of structured low-amplitude interference embedded in a bright beam \cite{Berry2018}; the mathematical details of the bright sidebands are lost in the asymptotic polynomial growth.

In \cite{Dennis2010}, the paraxial polynomial beam containing the trefoil knot was derived as the leading term in the expansion of a superposition of Laguerre-Gaussian beams with $1/w$ as the small parameter; in fact \cite{Dennis2011}, the same polynomial beam structure can be embedded in an envelope with \emph{any} profile, provided the width of the envelope is large compared to the knot structure (such as a Bessel beam with small $k_r$).
We will take this approach for the rest of this section, and only consider the polynomial part of (\ref{eq:milnor}), ignoring the Gaussian factor.

The form of the given polynomial in (\ref{eq:milnor}) does not agree dimensionally if $r$ has units of length. 
To make this explicit, all occurrences of $r$ in (\ref{eq:milnor}) should be replaced by $r/r_0$, for some reference length $r_0$.
Figure \ref{fig:Rescaling} (a) shows the six zeros of (\ref{eq:milnor}) as singularities of the phase, which occur at $r = 0.46 r_0$ and $2.19 r_0$ (at $\phi = 2\pi j/3$, $j=0,1,2$). 
We now denote the dimensionless length $k r_0$ by $s$; this parameter determines the size of the knot and, in the paraxial regime, $s\gg 1$, for which knots are large and elongated. 
The paraxial polynomial beam solution agreeing with (\ref{eq:milnor}) by $r\rightarrow s^{-1} k r$ is
\begin{align}
\label{eq:psi}
   \psi_{\text{3D}}^{\text{par}}=& 1-s^{-2} k^{2} r^2 - s^{-4} k^{4} r^4 +s^{-6} k^6 r^6 - 8 s^{-3} k^{3} r^3\text{e}^{3 \mathrm{i} \phi} \notag \\ 
&-2 \rmi \left(s^{-2}+4 s^{-4} k^{2} r^2 -9 s^{-6}k^4r^4  \right) k z  \notag \\
&+8 \left(s^{-4} +9 s^{-6} k^2 r^2 \right)  k^{2} z^2-48 \rmi s^{-6} k^{3} z^3, 
\end{align}
where the first line is independent of $z$, and corresponds to (\ref{eq:milnor}).

Paraxial beams satisfy a particular scaling rule, from the paraxial equation: all occurrences of $r$ and $z$ appear as the dimensionless parameters $R \equiv k r s^{-1}$ and $Z \equiv k z s^{-2}$; in a mathematical expression for a paraxial beam, there should be no other occurrences of $s$ and $k$.
Therefore, it is convenient to rewrite the paraxial polynomial knot beam (\ref{eq:psi}) as
\begin{align}
\psi_{\text{3D}}^{\text{par}}=& 1 - R^2 - R^4 + R^6 - 8  \:\!  \mathrm{e}^{3 \mathrm{i} \phi} R^3 - 2 \:\!  \mathrm{i}  Z - 8 \:\!  \mathrm{i}  R^2 Z \notag \\
& + 18 \:\!  \mathrm{i}  R^4 Z + 8 Z^2 - 72 R^2 Z^2 - 48 \:\!  \mathrm{i}  Z^3.
\label{eq:E-par}
\end{align}
It is straightforward to track the vortex lines as the complex zeros of the polynomial (\ref{eq:E-par}) using numerical root finder in a symbolic algebra package, or more complicated vortex tracking routines \cite{Taylor}.
The identification of the knot/link type for the simple vortex configurations we describe can be done by visual inspection, but, alternatively, one could implement more sophisticated algorithms \cite{Kauffman, Orlandini2007}.
The 3D-vortex configuration of (\ref{eq:E-par}) is shown in Figure \ref{fig:Rescaling} (b) in red: a trefoil knot is enclosed in the cylinder with diameter $d = 0.700 s^{-1}\lambda$ and height $h = 0.116 s^{-2}\lambda$; from the paraxial scaling, its transverse size is proportional to $s$, and longitudinal size to $s^2$.
Therefore our paraxial knot is much longer longitudinally than transversally.

It is very straightforward to perform a similar paraxial analysis for the other $m,n$-torus knots as described in section \ref{sec:Introduction}, but with the trefoil profile (\ref{eq:milnor}), rewritten in scaled coordinates (\ref{eq:E-par}), replaced with the appropriate section of a Milnor polynomial.
The results of the paraxial dimensions for several of these knots is given in in Table \ref{tab:scaling}.

\setlength{\tabcolsep}{3.6pt}
\begin{table}
\centering
\caption{
   \label{tab:scaling}\textbf{Paraxial scaling for different knots}. 
Transverse and longitudidinal sizes $d$ and $h$ for optical vortex $m,n$-torus knots and links in paraxial beams, generated as discussed in the Introduction. 
The main example in the paper is (2,3).}
\begin{tabular}{|c|ccccccc|} 
\hline
Knot type&(2,2)& (2,3)& (2,4) & (2,5) & (3,2) & (4,2) & (5,2) \\ 
\hline 
$d(10^{-3} s^{-1}\lambda)$ & 768&700 & 655 & 625 & 852 & 920 & 979 \\
$h(10^{-3} s^{-2}\lambda)$ &205& 116 & 79.0 & 58.7 & 173 & 152 & 113 \\ 
\hline
\end{tabular}
\end{table}

Although the longitudinal component is negligible in the paraxial regime, it is not zero.
To leading order in $k^{-1}$, $E_z = -k^{-1} \nabla\cdot (E_+ \widehat{\boldsymbol{e}}_+ + E_- \widehat{\boldsymbol{e}}_-)$; with the transverse polarization fixed as $\widehat{\boldsymbol{e}}_-$, this gives the longitudinal polarization as
\begin{align}
E_z^{\text{par}}=& - \sqrt{2} \:\!  \mathrm{i} s^{-1} R \mathrm{e}^{-\rmi \phi}  \left[ 1 + 2 R^2 - 3 R^4 + 24 R \:\!  \mathrm{e}^{3 \mathrm{i} \phi} \right.\notag \\ 
&\left.  +4\mathrm{i} Z\left(2 - 9  R^2\right) +72 Z^2 + O(s^{-2}) \right].
\label{eq:Ezpar}
\end{align}
The field $E_z^{\text{par}}$ satisfies the paraxial wave equation; ignoring the prefactor and terms in $s^{-1}$ and higher, it is given purely in terms of $R$ and $Z$ and hence has the same paraxial scaling as the transverse knot.  
Owing to spin-orbit conversion \cite{Bliokh2015}, it has a negative vortex on the beam axis; otherwise the quartic polynomial in $R$ has a knotted nodal structure (shown in Figure \ref{fig:Rescaling} (b)), however with inner vortex at $R=0.042$, almost at the origin. Such a knot cannot practically be measured; in the next section, we will consider the exact fields outside the paraxial approximation, where the vortices are separated by distances of order $s^{-1}$ (and when in fact $E_z$ is no longer negligible).

If, on the other hand, the initial transverse polarization was chosen to be $\widehat{\bs{e}}_+$, the resulting longitudinal field is
\begin{align}
E_z^{\text{+,par}} =& - \sqrt{2} \:\!  \mathrm{i} s^{-1} R \mathrm{e}^{\rmi \phi}  \left[ 1 + 2 R^2 - 3 R^4  \right. \notag \\ 
&\left. + 4 \mathrm{i} Z \left(2 - 9 R^2\right)  +72 Z^2 + O(s^{-2}) \right].
\label{eq:Ez+par}
\end{align}
Unlike $E_z^{\text{par}}$, the longitudinal component here is an eigenfunction of OAM with azimuthal quantum number unity; therefore there can be no other vortex structure in the longitudinal component.
When the OAM of the amplitude structure has the same sign as the polarization helicity, there in no extra knot-like interference structures in the longitudinal component.
This justifies our main investigation in the topologically more interesting case of imprinting the knot with positive OAM in a field of left-handed polarization. 
Apart from missing the extra azimuthal term to OAM, the radial dependence of $E_z^{\text{+,par}}$ is the same as $E_z^{\text{par}}$. 
More details are given in the Appendix.

Assuming the monochromatic magnetic field has an analogous form $\bs{B}\rme^{\rmi (k z- \omega t)}$, Maxwell's equations can easily be solved to show, within the paraxial approximation (ignoring corrections of order $s^{-1}$ and higher), that $\bs{B} \approx \mathrm{i} c^{-1} \bs{E}$, so, paraxially, the nodal knots in the magnetic components exactly agree with their counterparts in the electric field, as expected \cite{Berry2004c}.

We have therefore shown that in paraxial fields, the aspect ratio of a vortex knot in a transverse component scales with its transverse size, and that there is a knotted vortex structure in the negligible longitudinal component too.
From the scaling laws in Table \ref{tab:scaling}, getting a unit aspect ratio trefoil knot suggests bringing $s$ down to about $s_1 \approx 700/116 \approx 6$, which is outside the paraxial regime.
As $s$ approaches unity, we need to include the $s^{-1}$-dependent terms in the components of $\bs{E}$ and $\bs{B}$ \cite{Lax1975}, suggesting the vortex lines in each of the components behave differently.
We investigate this further in the next section.

\section{Exact analysis of nonparaxial polarization knot bundle}\label{sec:nonparaxial}

A polynomial solution of the paraxial equation can be made to solve the reduced Helmholtz equation by replacing monomials in $z$ by reverse Bessel polynomials \cite{Dennis2011}, which does not affect the initial condition in the focal plane $z = 0$.
Doing this to our main electric field component multiplying $\widehat{\bs{e}}_-$, we get
\begin{align}
E_-^{\text{np}} = E_-^{\text{par}} + s^{-2}\left[8 \:\! \mathrm{i} \left(1 + 18 s^{-2} - 9 R^2\right) Z +144 Z^2  \right],
\label{eq:E_-nonpar}
\end{align}
where, as in the previous section, we use $R = k r s^{-1}$, $Z = k z s^{-2}$.

Similarly, using Maxwell's equations and enforcing the polynomial beam ansatz, we find the following full forms for the components of the nonparaxial longitudinal electric field, and the magnetic field, correctly incorporating corrections in $s^{-1}$ and higher powers, 
\begin{align}
E_z^{\text{np}}  =& E_z^{\text{par}} +  4 \sqrt{2} \:\!  \mathrm{i} R \text{e}^{-\mathrm{i} \phi}   s^{-3} \left[2 + 54 s^{-2} -9 R^2 - 54 \:\!  \mathrm{i} Z \right], \label{eq:E_znonpar} \\
B_+^{\text{np}}  = &- 4 \:\!  \mathrm{i}c^{-1} R^2\text{e}^{-2  \:\!\mathrm{i} \phi} s^{-2} \left[  1 + 18 s^{-2} -3 R^2 -18\mathrm{i} Z \right. \notag \\
&+ \left. 24\:\!R^{-1} \text{e}^{3 \:\!  \mathrm{i} \phi} \right], 
\label{eq:B_+nonpar} \\
B_-^{\text{np}}=& \mathrm{i} c^{-1} \left\{E_-^{\text{par}} + 8 s^{-2} \left[ -36 s^{-4} -  s^{-2} + 9  s^{-2} R^2 \right. \right. \notag \\
&+ \left. \left.\mathrm{i} \left(1 + 36s^{-2}- 9 R^2  \right) Z+18 Z^2   \right] \right\},
 \label{eq:B_-nonpar}\\
B_z^{\text{np}} = &\mathrm{i} c^{-1} \left\{ E_z^{\text{par}} -  4 \sqrt{2} \:\!  \mathrm{i} R \text{e}^{-\mathrm{i} \phi}   s^{-3} \left[2 +36s^{-2} - 9 R^2  \right. \right. \notag \\
& -  \left. \left. 18  \:\!  \mathrm{i} Z \right] \right\}.
 \label{eq:B_znonpar}
\end{align}
The extra terms $\left[ \bullet \right] $ imply the scaling laws derived in the previous section do not hold for small values of $s$, for which the beam approaches the nonparaxial regime.
Tracking the nodal lines in each components can again be done easily, and the nodal knots in each component change shape when $s \lesssim 10$.

The exact forms (\ref{eq:E_-nonpar})--(\ref{eq:B_znonpar}) show how the exact optical energy is distributed around the focal volume in each component.
As $s \ll 1$, the transverse knot structure becomes subwavelength, which may be considered extremely nonparaxial, within the focal volume.
In this extreme nonparaxial regime, around the knot structure, $E_-^{\text{np}} \appropto s^{-4}$ whilst $E_z^{\text{np}} \appropto s^{-5}$.
Analogously, $B_+^{\text{np}} \appropto s^{-4}$, $B_-^{\text{np}} \appropto s^{-6}$, $B_z^{\text{np}} \appropto s^{-5}$.
As we will see in our more detailed analysis, the knots in each component dissolve away by reconnection when $s$ is close to unity and the field in the longitudinal components becomes large.
This phenomenon was already observed for optical vortex knots in \cite{Berry2001b, Dennis2009}.

\begin{figure}[h]
\centering
\includegraphics[scale=1]{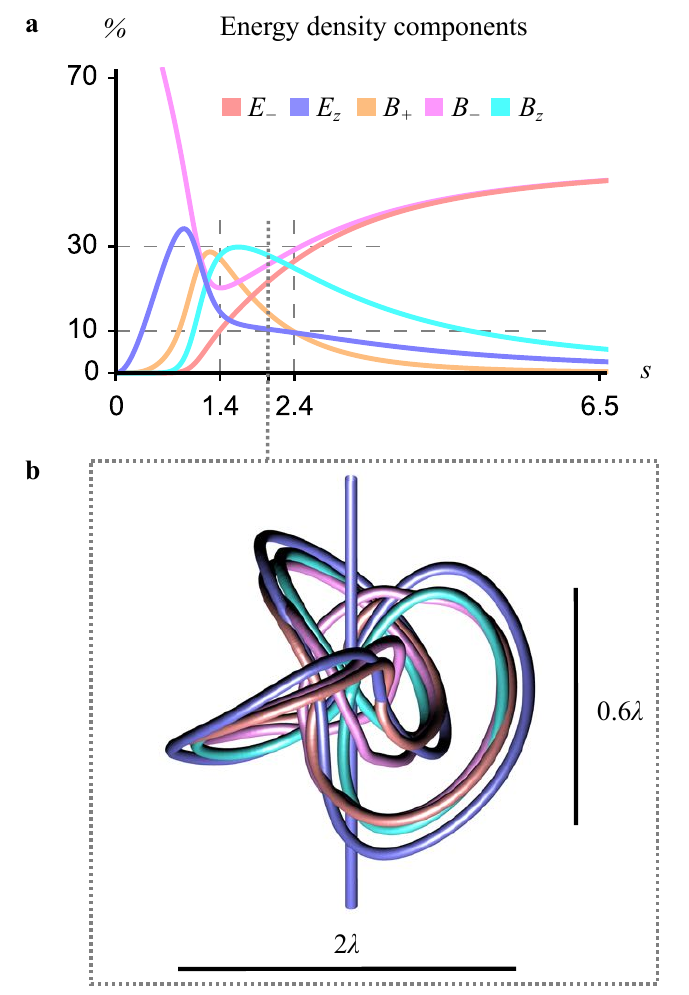}
\caption{ 
   \textbf{Singular knot bundle when all components have similar focal energy}.
   (a) Integrated energy density in the focal disk $z = 0$, $R \le  2.19$ for each of the components of the electromagnetic field as color coded.
   $E_+$ is exactly zero in the polynomial representation of the field.
   The dashed lines indicate that for $1.4 \le s \le 2.4$, each of the components (except $E_+$) has an energy between $10\%$ and $30\%$ of the total in the disk.
   (b) Singular polarization bundle.
   Nodal lines in the components of $E_-, E_z, B_-, B_z$ are all entwined, each in the form of a knotted line with a similar size, which is approximately within a cylinder of diamater $2.02 \lambda$ and total height $0.648 \lambda$.
   The longitudinal components also exhibit an axial vortex.
    \label{fig:Bundle} }
\end{figure}

As discussed in the Introduction, there is a range of $s$ where the energy in the neighborhood of the focal point is similar in the longitudinal and transverse components.
We determine the \textit{optimal} range of $s$ for this crossover by calculating the total energy in each component in the focal plane $z=0$, in the disk of radius $R = 2.19$ (the majority of energy, of course, being outside this).
We plot the energy in this disk in each component as a function of $s$ in Figure \ref{fig:Bundle} (a).
For $1.4 \lesssim s \lesssim 2.4$ the focal energy in all four components $E_-, B_-, E_z, B_z$ is of similar order ($10$--$30 \%$ each). 
This range of $s$ gives a crossover regime in which all the different vortex knots identified in the paraxial regime exist and occupy a similar volume, but are sufficiently nonparaxial for the spatial conformations of the knots to be different.
The resulting system of overlapping vortex knots in the different components of the electromagnetic fields forms a \emph{singular knot bundle}.
Figure \ref{fig:Bundle} (b) shows such remarkably complicated vectorial electromagnetic field distribution for the optimal value $s=2.05$, where the knotted nodal lines in each of the electromagnetic field components are entwined without coinciding.
Identifying and understanding this structure is the main result of this work.

\begin{figure}[h]
\centering
\includegraphics[width=\columnwidth]{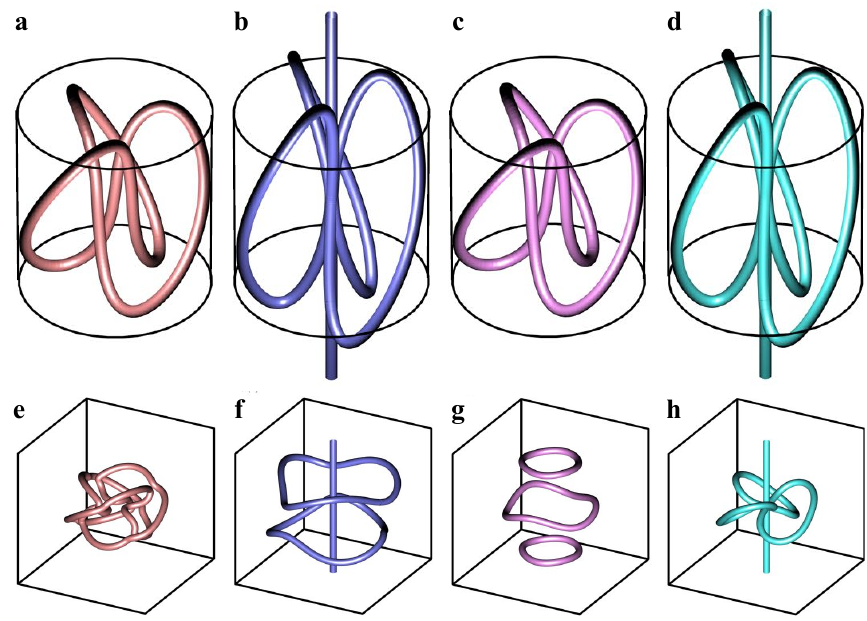}
\caption{
   \textbf{Nonparaxial vortex knots in different components of $\bs{E}$ and $\bs{B}$ fields}. 
(a) The trefoil knot in  $E_-^{\mathrm{np}}$ (a) has unit aspect ratio when $s_1 = 5.75$; the vortex lines in the other components also display a trefoil knot: (b) $E_z^{\mathrm{np}}$; (c) $B_-^{\mathrm{np}}$ (which resembles (a)); (d) $B_z^{\mathrm{np}}$ (which resembles (b)).
The cylinder $d_1= h_1 = 4.01 \lambda$ is shown around each vortex configuration for comparison. 
(e)--(h) The corresponding vortex configurations when $s = 1.00$, drawn in a cube $\lambda^3$.  
(a) The knot in $E_-^{\mathrm{np}}$ corresponds to a value of $s$ just above the value where its lines reconnect, which are bounded by a cylinder of subwavelength volume $V_c \approx 0.15 \lambda^3$; (f) the vortex lines in $E_z^{\mathrm{np}}$ have already reconnected to form two approximately coaxial rings, and an axial vortex; (g) vortex lines $B_-^{\mathrm{np}}$ have reconnected to give three approximately coaxial rings; (h) vortex lines in $B_z^{\mathrm{np}}$ is still a trefoil knot, threaded by a vortex line.
}
\label{fig:Unit}
\end{figure}

The singular knot bundle of Figure \ref{fig:Bundle} (b) has an aspect ratio of $0.32$, smaller than unity.
By numerically tracking the vortices for the nonparaxial field (\ref{eq:E_-nonpar}) for $s$ close to $6$ the paraxially-estimated value for the unit aspect ratio, we find a knot of aspect ratio unity when $s = s_1 \equiv 5.75$, for which $d = h = 4.01 \lambda$, shown in Figure \ref{fig:Unit} (a).
Thus, the previously published knotting scheme \cite{Dennis2010}, appropriately rescaled, does give a knot of unit aspect ratio but only when scaled to a few optical wavelengths across.
The vortices in $E_z^{\mathrm{np}}, B_-^{\mathrm{np}}$ and $B_z^{\mathrm{np}}$ are shown in Figure \ref{fig:Unit} (b, c, d); as with the knot in $E_-^{\mathrm{np}}$, these all closely resemble their paraxial counterparts (e.g.~the transverse components, and longitudinal components, have zeros very close, with an axial longitudinal vortex). 
In particular, the small $s$ corrections to the longitudinal components are sufficient to ensure the distances between the inner vortices of the longitudinal trefoil knots become comparable with their transverse size (with the vortex up the axis), but not large enough to split the conformations of the knots in the electric and magnetic fields, which are effectively coincident.

Throughout this section, we have been considering as a thought experiment the propagation of the electromagnetic field in the neighborhood of the focal point, given that the $E_-$ component is given by (\ref{eq:E-par}) and $E_+ \equiv 0$.
If $s$ is made so small that the transverse knot structure is subwavelength, we would not be surprised if nonparaxial effects are sufficiently large to disrupt the knot topology, causing reconnections that dissolve the knot in both this and the other components \cite{Berry2012}.
The case of $s = 1.00$ is shown in Figure \ref{fig:Unit} (e)-(h); the knot exists in the $E_-^{\mathrm{np}}$ and $B_z^{\mathrm{np}}$ components, but have already reconnected away in the $B_-^{\mathrm{np}}$and $E_z^{\mathrm{np}}$ components.
The transverse vortex knot in $E_-^{\mathrm{np}}$ (Figure \ref{fig:Unit} (e)) here is so small it fits within a cubic wavelength, proving that subwavelength nodal knots are supported by Maxwell's equations.
The reason for the radially symmetric behavior in both $E_z$ and $B_-$ can be extrapolated by Maxwell's equations.
The details of the proof are in the Appendix.

\begin{figure}[h]
\centering
\includegraphics[scale=1]{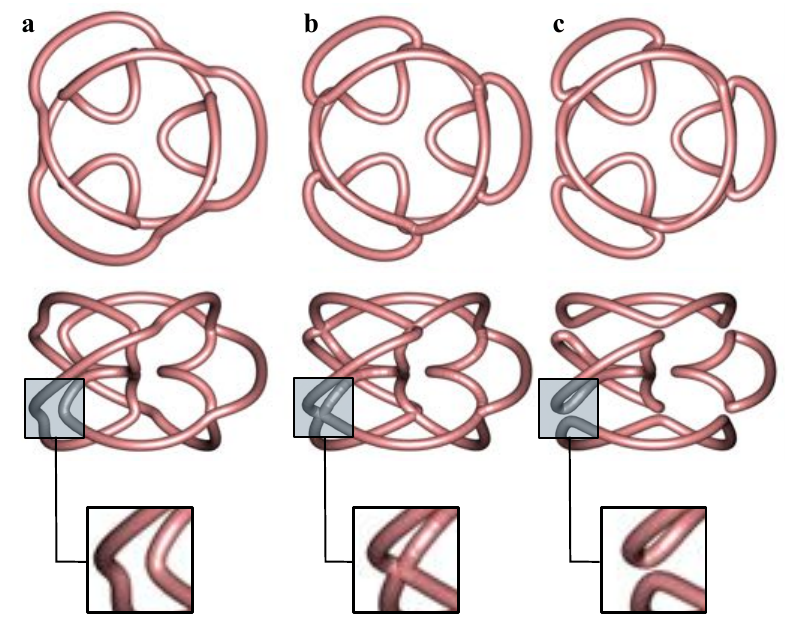}
\caption{ 
   \textbf{Reconnection events destroying vortex knot in $E_-$ component}. 
   The parameter $s$ is decreased from (a) $s = 1.000$, where there is a knotted vortex line, through the critical value of $s_c=0.965$ where the vortices touch at six points. At lower values of $s$, such as (c) $s = 0.950$, the vortex configuration is five unlinked rings.
   In each frame, there is an axial view, a side view and an inset showing the vortex geometry. 
    \label{fig:Reconnection} }
\end{figure}

The details of how the knot in $E_-^{\mathrm{np}}$ dissolves away by reconnections when $s < 1$ is illustrated in Figure \ref{fig:Reconnection}.
As the parameter $s$ reduces below unity, the vortex lines in the principal $E_-^{\mathrm{np}}$ component (Figure \ref{fig:Reconnection} (a)) approach each other at six points on the knot, and reconnect at the critical value $s_c = 0.965$.
After the topology has changed, the vortex configuration consists of five loops, one above and one below the focal plane, and three symmetrically arranged within the focal plane (Figure \ref{fig:Reconnection} (c)).
The volume of a bounding cylinder around the knot at $s_c$, immediately prior to its destruction, is $V_c= 0.15 \lambda^3$.
The sequence of events dissolving the topology of the knots in the other components is reported in the Appendix, showing similar behaviours to $E_-$.

\setlength{\tabcolsep}{2.55pt}
\begin{table}[htbp]
\centering
\caption{
   \label{tab:nonpar}\textbf{Nonparaxial torus knot bundles}. Values $s_1$ for which the aspect ratio of the knot in $E_-$ is $1$ and the correspondent dimensions $d_1=h_1$. 
   Limiting values $s_c$ for which each knot and link dissolves and correspondent volume $V_c$ of the cylinder that encloses the nodal lines. Knot type in $E_z$. 
   The case study here is (2,3).}
\begin{tabular}{|c|ccccccc|}  
\hline
$E_-$ knot type&(2,2)& (2,3)& (2,4) & (2,5) & (3,2) & (4,2) & (5,2) \\ 
\hline
$s_1$&$3.59$&$5.75$&$7.90$&$10.2$&$4.85$&$5.80$&$8.40$\\
$d_1(\lambda)$&$2.76$&$4.01$&$5.18$&$6.40$&$3.98$&$5.35$&$8.22$\\
$s_c$&$0.36$&$0.97$&$3.10$&$4.50$&$0.71$&$0.88$&$3.10$\\
$V_{c}(\lambda^3)$&$0.013$&$0.151$&$3.37$&$9.62$&$0.080$&$0.338$&$11.1$\\
$E_z$ knot type&unknot&(2,3)& (2,4) & (2,5) &(2,2)& (3,2) & (4,2)\\
\hline
\end{tabular}
\end{table}

The topological events described in this section are are not unique to the trefoil knot. 
The values of $s$, $d$, $h$ for unit aspect ratio and reconnection events are reported in Table \ref{tab:nonpar} for several different torus knots and links.
This characterizes general optical vortex knots in the nonparaxial regime. 
We tracked the vortex lines of $E_z$ for other knots and we reported their knot type in Table \ref{tab:nonpar}.
In the Appendix, we present similar but more general argument than in the main text anticipating the relationship between the knots in the various components of the polynomial beam electromagnetic field for initial conditions analogous to (\ref{eq:milnor}) for different knots.

The beams described in this section could be implemented in the laboratory by tightly focusing a paraxial beam  or by constructing nanoscale resolved holograms (both equivalent to a small value of $s$  which brings the paraxial optical beam of \cite{Dennis2010} to the nonparaxial regime). 
However, the physical challenge of polynomial beams relies in the fact that the bulk of the energy in these beams is concentrated away from the knotted structure, and the detection of nodal knots in subwavelength volumes of extremely low intensity is a considerable challenge for experiment. 
Hence, a less complicated, realistic experimental design is required to potentially measure a polarization knot bundle. 
In the next section, we propose an experimental design by which the singular knot bundle presented here could be observed in a physical system.

\section{Experiment design to measure the bundle in tightly focused beams}
\label{sec:experiment}

Here we propose an experimental scheme that gives a singular knot bundle within the focal volume of a microscope objective of high numerical aperture. 
The general topological properties of the knotted fields described in the previous section are not unique to polynomial beams, and here we demonstrate the existence of the polarization knot bundle numerically in tightly focused beams, an approach that relies on slightly different assumptions than the exact polynomial beam-based analysis of the previous section.
The analytical treatment from the previous section backs up the simulations, even though the shape of the knots in the bundle are slightly different, indicating that the phenomenon we describe is general to nonparaxial beams of diverse nature.
We predict that the structure we present here could be measured with the polarimetry techniques reported in \cite{Bauer2013, Miles2015} if they were optimized for the detection of low-amplitude optical fields.

\begin{figure}[h]
\centering
\includegraphics[scale=1]{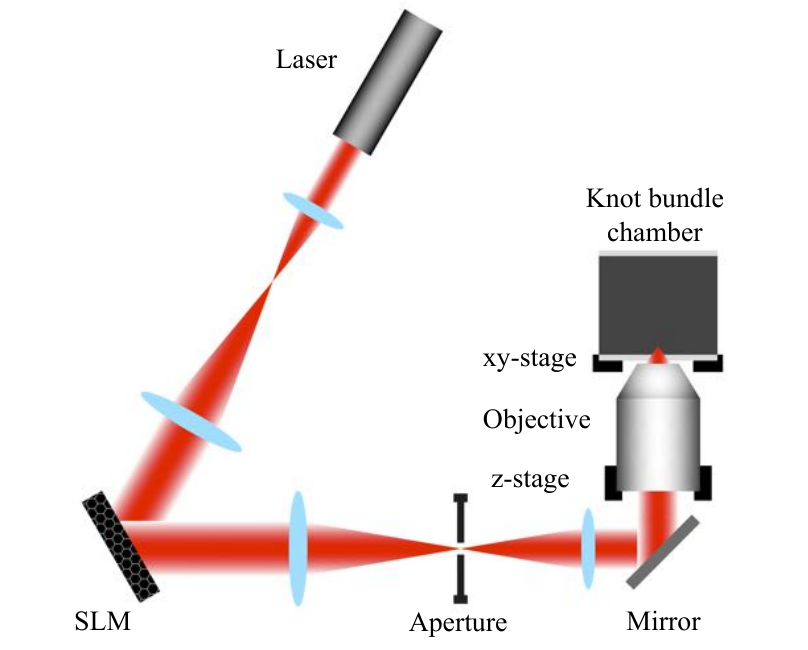}
\caption{
   \textbf{Experimental scheme}. 
Experimental setup adapted from \cite{Leach2004, Leach2005, Dennis2010} of a tightly focused beam for the creation of the bundle of polarization singularities. 
The numerical aperture of the microscope objective is $\gamma \geq 0.90$, the radius of the back focal plane  is $a = 7000 \lambda$. 
In practical experiments a uniformly circularly polarized beam acquires a a weak transverse component of opposite handedness and a longitudinal component. 
When the hologram (\ref{eq:hologram}) is generated in the spatial light modulator (SLM), the beam in the focal volume has knotted nodal lines in each component of the electric and magnetic fields .}
\label{fig:ExperimentSetup}
\end{figure}

Nonparaxiality is achieved in this proposed experiment by including a high numerical aperture microscope objective in the setup from \cite{Leach2004, Leach2005, Dennis2010}, giving raise to tightly focused beams.
The experimental configuration is shown in Figure \ref{fig:ExperimentSetup}. 
For practical purposes, the hologram plate is specified in Fourier space; the resulting structured beam has to be circularly polarized (along $\widehat{\bs{e}}_-$) before being sent to the back focal plane of the microscope objective (the polarization setup is not included in the scheme for simplicity). 
When such beams with a well-defined transverse polarization state are tightly focused, spin-orbit effects result in a nonzero field in the orthogonal transverse polarization component $\widehat{\bs{e}}_+$ \cite{Leuchs2003, Zhao2007}, as well as the principal components $E_-$, the longitudinal field $E_z$ and their magnetic counterparts.
This redistribution of the energy between the different components is due to the aperture-lens system and is more significant for larger tight focusing.

We found that it should by possible to create a singular knot bundle in the focal volume of a microscope objective of high numerical aperture ($\gamma \geq 0.9$, assuming index of refraction $1$) and radius of the back focal plane $a = 7000 \lambda$.
Our approach to design the hologram function consists in calculating the Fourier transform of (\ref{eq:milnor}) and discarding the terms that are not necessary for the knot bundle to appear under tight focusing.
The numerically optimized Fourier hologram takes the form (in cylindrical coordinates $\rho$ and $\varphi$, in units of aperture $a$ and illuminating Gaussian of waist width $w_0$) 
\begin{equation}
   E_{-}(\rho,\varphi;w_0) =\mathrm{e}^{ -\rho^2 /(2 w_0^2)} \left(2 + 3\, \rmi \, \rho^3 S^{-3} \mathrm{e}^{ 3 \, \rmi \,\varphi} + 3 \rho^4 S^{-4} \right).
\label{eq:hologram}
\end{equation}

We propagated the function in (\ref{eq:hologram}) in the focal volume by means of vector diffraction theory \cite{RW1959, Novotny}.
Both the paraxial and the extreme nonparaxial regimes do not give the knot bundle (as in the previous section); an appropriate balance between them is required in order to create the bundle regime.
Such an overcrossing regime can be found by the careful manipulation of the focusing parameters $\gamma$, $w_0$ and $S$.
In analogy to \cite{Dennis2010, Padgett2011}, the tightly focused experimental scheme depends strongly on the waist of the illuminating Gaussian envelope $w_0$ incident on the backfocal plane of the microscope objective, which plays an important role in determining the topology of the beam.
In fact, small $w_0$ implies that the real space beam is enveloped in a wide Gaussian; this regime clearly resembles the polynomial beam propagation that we already stated is not practical. 
On the other hand, large $w_0$ overfills the aperture, creating additional diffractive rings in the focal plane, around a focal volume that is too small to contain the knotted structure (we saw in the previous section that the knot bundle cannot be made much smaller than $\lambda^3$), and tight wavefront curvature from small $1/w_0$ might further affect the knot topology, as found in the paraxial regime \cite{Dennis2010, Padgett2011}.
The transverse size of the knot is controlled by the parameter $S$, which scales the input beam by analogy with the polynomial beams (a direct scaling of the focal field through the  parameter $s$ would require more advanced methods to structure vector fields).

\begin{figure}[t]
\centering
\includegraphics[scale=1]{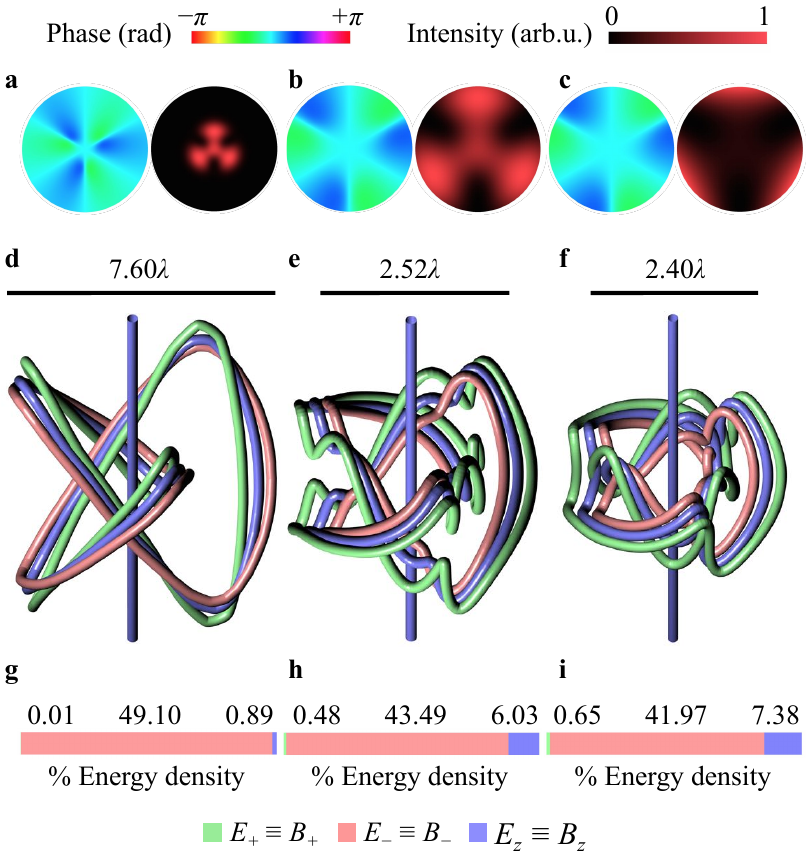}
\caption{
   \textbf{Numerical simulations}. 
The numerical propagation of the hologram plate (\ref{eq:hologram}) by Richards-Wolf vector diffraction theory gives the singular knot bundle, with focusing parameters $\gamma=0.9$, $S=0.27$, $w_0= 0.62 S$ (a, d, g); $\gamma= 0.95$, $S=0.75$, $w_0= 0.65 S$ (b, e, h); $\gamma= 0.95$, $S=0.90$, $w_0= 0.90 S$ (c, f, i). 
The beam's nodal lines in $\boldsymbol{E}$ and $\boldsymbol{B}$ are coincident, as well as their energy density distribution. 
(d) The focal volume has a trefoil knot  in each component of the electric and magnetic fields: $E_-$, $B_-$ (red); $E_+$, $B+$ (green) and $E_z$, $B_z$ (blue). 
(e) The trefoil knots are deformed and smaller for more tightly focused beams. 
(f) Each of the knots in $E_+$ and $E_z$ have dissolved into two separate loops because the beam is too nonparaxial, $E_-$ is still trefoil knotted. 
(g--i) The energy transferred from $E_-$ to $E_+$ and $E_z$ is larger for increasing tight focusing (equivalent distributions are found for the magnetic components).}
\label{fig:ExperimentKnots}
\end{figure}

By the analysis of the vortex lines of each component of the beam we found that the correct balance of the parameters is given by not overfilling waist of the Gaussian $w_0$, which gives an effectively smaller numerical aperture with respect to that of the microscope objective, but, more importantly, a larger focal volume in which to embed the knot bundle.
The holograms given by (\ref{eq:hologram}) for different sets of parameters are plotted in Figure  \ref{fig:ExperimentKnots} (a)--(c). 
Such functions do not contain any phase singularity and, as mentioned, their paraxially propagating beams do not embed knotted vortex lines.
This is because our optimized hologram (\ref{eq:hologram}) uses the lens-aperture system to shape the beam's interference pattern containing knotted structures. 

When $\gamma=0.90$, $S=0.27$ and $w_0= 0.62 S$ the singular knot bundle of Figure \ref{fig:ExperimentKnots} (d) appears in the focal volume.
The numerically propagated beam contains a trefoil knot in $E_-$ (red) and $E_z$ (blue).
From the spin-to-orbit effects of the input beam in the lens system, the nodal lines of the extra component $E_+$ are also trefoil knotted, threaded by an axial vortex (shown in the figure in green).
Three trefoil knots appear in all the components of the magnetic field at the same position, due to the dual properties of tightly focused circularly polarized beams. 
The knot in $E_+$ and the mutual linear dependence of the $\boldsymbol{E}$ and $\boldsymbol{B}$ fields are the major differences from the polynomial beams predictions. 
The differences from the polynomial beams arise physically from the fact that the lens and the aperture system inevitably affect the input beam, and the Gaussian envelope in the focal volume is too small to behave as a polynomial beam.
By evaluating the electromagnetic energy density in a disk of radius $R=3.80 \lambda$ (see previous section) we found that the energy transferred to the other components is very low, reported in Figure \ref{fig:ExperimentKnots} (g).
Specifically, only $0.01 \% $ goes to $E_+$ and $0.89 \% $ to $E_z$ (equivalently in $\boldsymbol{B}$).

In order to generate an energy distribution that could potentially be measured in an experiment, we tightly focus the beam more in order to create the bundle regime and without destroying the topology of the knotted lines. 
We do so by increasing the focusing parameters $\gamma= 0.95$, $S=0.75$ and $w_0= 0.65 S$.
As a result, the knot bundle of Figure \ref{fig:ExperimentKnots} (e) and the energy distribution (h) appear.
This knot bundle shows three deformed trefoil knots of smaller size compared to the previous case, which are a manifestation of the nonparaxial effects.
On the other hand, the energy is larger in $E_+$ and $E_z$ which are $0.48 \%$ and $6.03 \%$ in a radius $R=1.26 \lambda$.

A further optimization of the energy cannot be achieved because the knot bundle starts to dissolve in $E_+$ and $E_z$ when the beam is too tightly focused ($\gamma= 0.95$, $S=0.90$, $w_0= 0.90 S$), whilst the trefoil knot is still present in $E_-$, as shown in Figure \ref{fig:ExperimentKnots} (f).
This choice of parameters has $0.65 \%$ of energy  in $E_+$ and $7.38 \%$ in $E_z$ in the disk $R=1.20 \lambda$ (Figure  (i)).
This shows that, consistently with the theoretical results of the previous section, the knot bundle does not appear in an extreme nonparaxial regime, for which its volume is smaller than $\lambda^3$.
However, the way the trefoil knots dissolve is different from the previous section: both the knots in $E_+$ and $E_z$ split only into two loops here.
This is the third major difference from the knots embedded in polynomial beams.

The numerical experiment  we designed can be easily extended to other knots and links. 
The topological behavior of the nodal lines of the different components can be predicted with polynomial beams, and the bundle regime can be approached in a high numerical aperture system by choosing the focusing parameters appropriately.
As already stated in the previous section, a good balance between all the components of the field seems to be a requirement for the successful creation of knotted nodal lines in each component of the electromagnetic field. 
Our numerical results show that the singular knot bundle seems to be a general phenomenon of nonparaxial knotted beams in which spin to orbit effects are dominant. 

\section{Discussion}
\label{sec:discussion}

We have shown how nonparaxiality and tight focusing affects the optical vortex torus knots of \cite{Dennis2010}. 
At small sizes, polarization effects become important, revealing a new sort of knotted optical field enclosing a complicated structure we call a singular knot bundle. 
In order to establish the generality of the knot bundle, we investigated two different kinds of nonparaxial electromagnetic fields.
Polynomial beams \cite{Dennis2011} give an analytic expression in the neighborhood of the knot, thereby revealing general behavior of the optical vortex topology, whereas numerically propagated tightly focused Laguerre-Gaussian beams \cite{RW1959, Novotny} describe a more physically accessible situation realizable in experiments.
Although we mainly focused on the specific example of the trefoil knot, in the Appendix we outline a general argument establishing bundles of torus knotted polarization singularities as general in appropriate superpositions of nonparaxial, circular polarized beams.
The bundle is a topological object, where the nodal lines in each component of the electric and magnetic fields are simultaneously knotted in a similar volume (the knots are not necessarily of the same type).
Its presence is hidden in the paraxial regime in which electric and magnetic vortex knots collapse  and some components are negligible, so the knot bundle effectively behaves like one single vortex line.
We identified a particular regime---the bundle regime---in which all the different knots become entities in their own right because the electromagnetic energy is distributed equivalently in all the components.
We expect this phenomenon to extend to other, more complicated knots (including non-torus knots such as the figure-eight knot), suggesting this nonparaxial polarization structure adds to the menagerie of knotted electromagnetic fields.
In particular, other knotted physical quantities related to knotted light are the time dependent electromagnetic field lines and their related null lines \cite{Irvine2008, Kedia2013} and the static optical vortex lines of complex scalar fields \cite{Berry2001a, Berry2001b, Leach2004, Dennis2010} (the latter construction was extended to the longitudinal component of nonparaxial beams in \cite{Maucher2018}).

Instead of tracking the nodal lines of each component of the electric and magnetic field in the canonical helical basis, our topological tools could be used to account for singularities that are independent of the choice of basis, such as C\textsuperscript{T} lines and L\textsuperscript{T} lines \cite{Nye1987}.
It is known that in the perfect paraxial regime C and C\textsuperscript{T} lines sit together and as they move to the nonparaxial regime they form clusters of polarization singularities \cite{Berry2004c}.
In the polynomial beams, we assumed $E_+$ to be exactly zero everywhere and the longitudinal component to be non-negligible; with these assumptions the nodal lines of $E_-$ are L\textsuperscript{T} lines (the electric field is only longitudinal along those lines) and the nodal lines of $E_z$ are C\textsuperscript{T} lines (the electric field is purely left-handed circular along those lines). 
On the other hand, the magnetic field has also a right-handed component; hence, its polarization singularities are not \textit{true} with respect to the 3D vector and the nodal lines of $B_+$ and $B_-$ are C lines (only the transverse vector is circularly polarized) and the nodal lines of $B_z$ are points along which the field is purely transverse (mixture of right and left handed polarization states). 
The polarization on the optical axis is purely left-handed circular for both $\bs{E}$ and $\bs{B}$ since all the other components are null (it truly behaves like an optical vortex line).
However, in tightly focused beams the topology of C\textsuperscript{T} and L\textsuperscript{T} lines is not this obvious.
A preliminary analysis of the C\textsuperscript{T} lines of our numerically propagated fields do not show knots, but other interesting conformations such as loops and lines intertwining the optical axis.
On the other hand, the L\textsuperscript{T} lines do not present any topology of interest, mainly because the knotted condition is extremely hard to achieve for these type of singularities.
Overall our results indicate that the topology of tightly focused beams is very rich and much more could be revealed by the topological analysis of the singularities of other physical quantities.
Beam parameters could potentially be optimized to structure these various kinds of polarization singularities, such as the time averaged and instantaneous Poynting vector, optical field lines and the Riemann-Silberstein vector.

Our theoretical analysis revealed the presence of a bundle regime, for which the energy is approximately equally distributed between all the components of the beam; this range seems to be ideal not only from the point of view of topology but also that of experiments.
We proposed an experimental scheme to generate the bundle regime and we highlighted the difficulties to create the right balance between the experimental parameters for the generation of the singular knot bundle.
Our experimental setup  was adapted from that of previous paraxial optical vortex knot experiments \cite{Dennis2010}, with the difference that the beam is tightly focused by a microscope objective and its initial polarization state is circular.
As with all experiments shaping superoscillatory phenomena \cite{Berry2018}, 
the beam should be accurately structured at subwavelength scale and imperfections such as aberrations should be reduced to a minimum, and the low amplitude around the nodal structures needs to be resolved and distinguished from the CCD camera noise. 
Measuring the 3D structure of the 3D polarization field in the focal volume also presents an experimental challenge; this might be approached via 3D nano-tomography similar to \cite{Bauer2015}.
Our experimental design aims to be as close as possible to the experimental routines currently used in structured light where Laguerre-Gaussian modes seem to be the basis of preference. 
Nevertheless, polynomial beams could effectively represent superoscillatory interference structures close to the axis within a beam with a wide envelope, not only a Gaussian, but also, for instance Bessel beams \cite{Durnin1987a, Durnin1987b}, and alternative envelopes could be designed to fit a specific experimental setup; for example, by exploring Richards-Wolf propagation theory further, the inverse problem could be solved in a more sophisticated way to generate new holograms for knotted beams.
An alternative way to generate and detect these optical knotted fields could be via single photon measurements; the experimental methods of \cite{Romero2011, Tempone-Wiltshire2016} could be adapted to the nonparaxial regime.

The knotted structures we describe, for the $m,n$-torus knots, occur in superpositions of two OAM eigenstates $\ell=0$ and $\ell=n$, and knots in similar fields, at the single photon level, were previously used as a basis for spatial measures of quantum entanglement \cite{Romero2011}.
In nonparaxial propagation, azimuth-dependent effects (`orbital') becomes strongly coupled with polarization (`spin') \cite{Bliokh2015}. 
The singular knot bundle may therefore be considered as a particular macroscopic 3D manifestation of optical spin-orbit interaction, when the OAM has an opposite sign from the spin.
This kind of structured, tightly focused light might be used to induce circular dichroism into chiral and nonchiral structures in order to provide morphological information of certain nanoscale structures, such as proteins \cite{Kelly2005} , metamaterials \cite{Decker2007} and plasmonic systems \cite{Zambrana-Puyalto2014, Gorodetski2009}.

Our investigation of small optical vortex knots, originally motivated by imprinting optical knots into matter, has led to the discovery of the polarization bundle as a quite general phenomenon in knotted light fields where the aspect ratio of the knot is approximately unity.
The singular knot bundle's stability in various nonparaxial beams demonstrates its robustness to perturbations of diverse origins and is worth investigating further. 
It would be interesting to investigate how such 3D spatially varying polarization fields affect materials that reorient with respect to the polarization direction, such as azobenzene polymers \cite{White2009, Wang2012, Bin2015}.
Embedding the singular knot bundle into soft-materials or quantum physical systems might reveal new features of light-matter interaction and could be used to store topological states \cite{Pugatch2007}.

\section*{Funding Information}
This research was funded by the Leverhulme Trust Research Programme Grant No.~RP2013-K-009, SPOCK: Scientific Properties of Complex Knots. 
This work was carried out using the computational fa cilities of the Advanced Computing Research Centre, University of Bristol.

\section*{Acknowledgments}

The experiment design was inspired by discussions with Peter Banzer and Martin Neugebauer, whom we thank for guidance. 
We also thank Michael Berry, Benjamin Bode, Etienne Brasselet, Henkjan Gersen, Martin Gradhand, Eileen Otte, Alexander Taylor and Teuntje Tijssen for discussions.

\renewcommand{\thefigure}{A\arabic{figure}}
\setcounter{figure}{0}
\renewcommand{\theequation}{A\arabic{equation}}
\setcounter{equation}{0}

\section*{Appendix}

\begin{figure*}[t]
\centering
\includegraphics[scale=1]{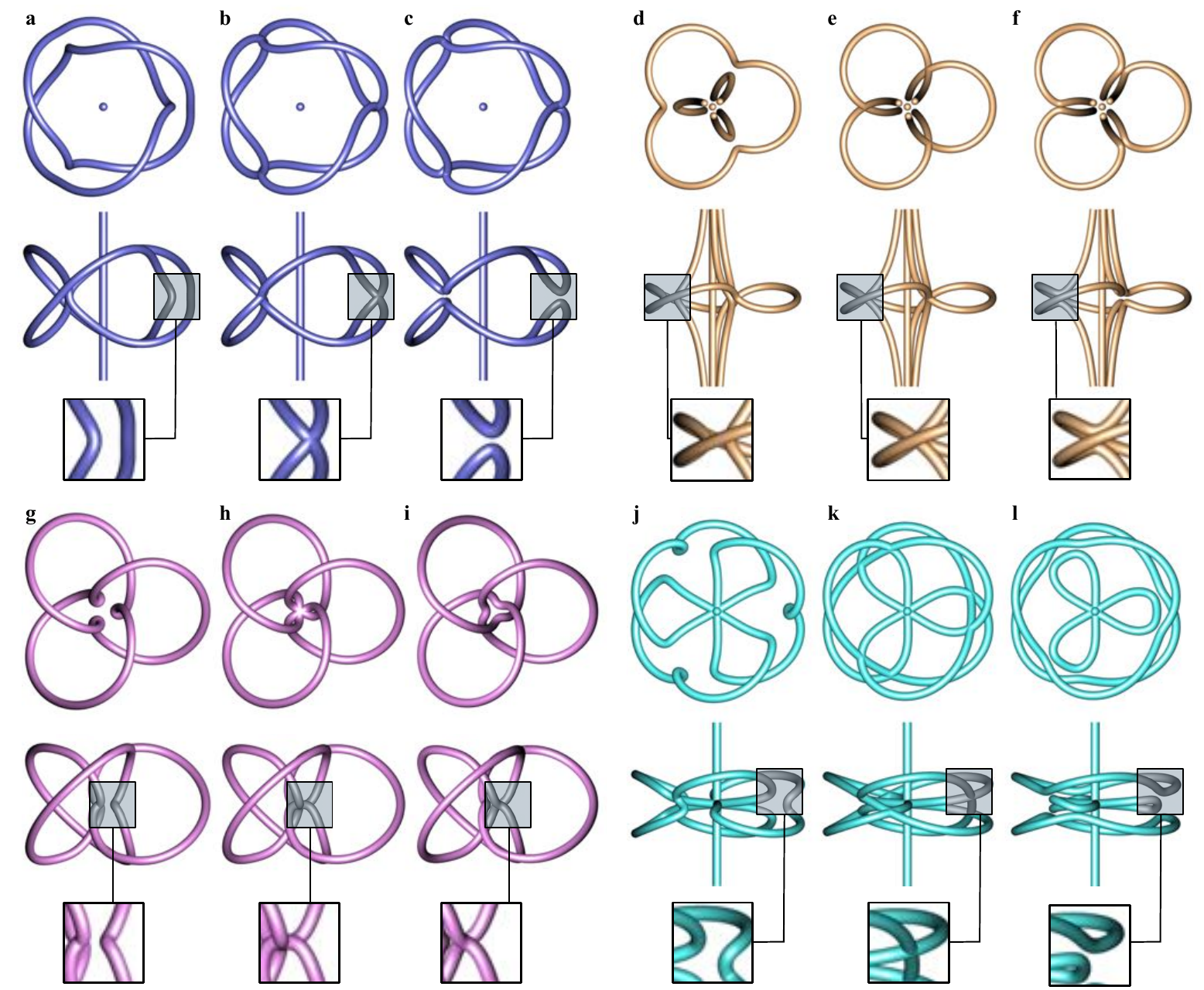}
\caption{
   \textbf{Reconnection events transforming the vortex lines in $\varphi$, $\overline{\vartheta}$, $\overline{\psi}$, $\overline{\varphi}$ for the trefoil knot (as in the main text).} 
   The parameter $s$ is decreased and topological reconnection occur at different values for the different components. 
   (a) The trefoil knot in $\varphi$ at $s=1.200$ (b) touches at three points at $s= 1.163$ and (c) splits into two circles showed at $s=1.115$. 
   The nodal lines of $\overline{\vartheta}$ are not knotted at any value of $s$, however (e) topological reconnections occur at $s=0.911$; the dynamic of the events is shown for (d) $s=0.920$ and (f) $s=0.900$.
   (g) The trefoil knot in $\overline{\psi}$ at $s=2.680$, (h) reconnects at the origin for $s=2.647$ and (i) transforms into the Borromean rings, shown at $s=2.620$. 
   (j) The trefoil structure in $\overline{\varphi}$ at $s=0.610$, (k) reconnects at $s=0.556$ and (l) splits into another trefoil-like knot and two loops shown at $s=0.535$.}
\label{fig:ReconnectionMORE}
\end{figure*}

In this section we show that the symmetries required by the knot bundle presented in the main text can be extrapolated directly from Maxwell's equations.
In particular, we consider the relationships between the different components of the electromagnetic field, assuming that one of the transverse circular components of the electric field is a Milnor-like function as discussed in the main text Introduction; this is straightforward for the torus knots.

We will follow the logic of sections 2 and 3, but with more general choices for the electromagnetic fields.
We assume the electric field has the form
\begin{equation}
   \bs{E} = \rme^{\rmi(k z - \omega t)}( \psi \widehat{\bs{e}}_{\mp} + \varphi \widehat{\bs{e}}_z),
   \label{eq:Escalars}
\end{equation}
where the knot-carrying initial condition is in the transverse component $\psi$ when $z=0$, which is either left-handed ($-$) or right-handed ($+$) circularly polarized.
The corresponding magnetic field (times a factor of $c$ to ensure our scalar fields have the same physical dimension), is 
\begin{equation}
   c\bs{B} = \rme^{\rmi(k z - \omega t)}( \overline{\psi} \widehat{\bs{e}}_{\mp} + \overline{\vartheta} \widehat{\bs{e}}_{\pm} + \overline{\varphi} \widehat{\bs{e}}_z),
   \label{eq:Bscalars}
\end{equation}
which, unlike the electric field, has in general nonzero components in both right-handed and left-handed transverse components.

In keeping with section 3 of the main text, we work in units of inverse optical wavenumber $k^{-1}$.
\[ X = s^{-1} k x, \, Y = s^{-1} k y, \, Z = s^{-2} k z, \]
where $s$ is a dimensionless scaling factor, with the paraxial regime corresponding to $s \gg 1$.
As we consider superposition of orbital angular momentum (OAM) eigenstates, it is further convenient to use complex helical coordinates $\zeta, \zeta^*$ in the transverse plane,
\begin{eqnarray} 
   \zeta = \tfrac{1}{\sqrt{2}}( X + \rmi Y ), &\quad & \zeta^* = \tfrac{1}{\sqrt{2}}( X - \rmi Y ) \label{eq:zetadef} \\
    X = \tfrac{1}{\sqrt{2}}( \zeta + \zeta^*), & \quad & Y = -\tfrac{\rmi}{\sqrt{2}}( \zeta - \zeta^*).
\end{eqnarray}

From these assumptions, we can now find the relationships between the five fields $\psi, \overline{\psi}, \varphi, \overline{\varphi}, \overline{\vartheta}$ through Maxwell's equations.
We assume $\psi$ is given as in section 3, from polynomial propagation of a polynomial initial condition at $Z = 0$.
We will establish how the knotted field $\psi$ affects the symmetries of the other components in the extreme regimes of large $s$ (paraxial) and limiting to $0$ (highly nonparaxial).
The symmetries of the function $\psi$ are transferred to the other components $\overline{\psi}, \varphi, \overline{\varphi}, \overline{\vartheta}$ in an overcrossing regime we call \textit{bundle regime} in the main text and 
under the right assumptions each component inherits the knotted topology from $\psi$. 
This is demonstrated in the following.

First, from $\nabla\cdot\bs{E}$, we have
\begin{equation} 
   \varphi = \rmi s^{-1} \left[ \left\{ \begin{smallmatrix} \partial_{\zeta} \\ \partial_{\zeta^*} \end{smallmatrix}\right\} \psi + s^{-1} \partial_Z \varphi\right],
   \label{eq:divE}
\end{equation}
where the upper alternative occurs for the $\bs{E}$-field being left handed polarized ($-$), and lower for right handed polarized ($+$).
Spin-orbit effects of course relate the helical coordinate derivatives to the helical polarizations.
This equation has the solution
\begin{align} 
   \varphi & =  - s \rme^{-\rmi s^2 Z} \int^Z \rme^{\rmi s^2 Z'} \left\{ \begin{smallmatrix} \partial_{\zeta} \\ \partial_{\zeta^*} \end{smallmatrix}\right\} \psi \rmd Z' \nonumber \\
   & =  \frac{\rmi}{s} \left[ \left\{ \begin{smallmatrix} \partial_{\zeta} \\ \partial_{\zeta^*} \end{smallmatrix}\right\} \psi - \rme^{-\rmi s^2 Z}\int^Z \rme^{\rmi s^2 Z'} \left\{ \begin{smallmatrix} \partial_{\zeta} \\ \partial_{\zeta^*} \end{smallmatrix}\right\}\partial_Z \psi \rmd Z'\right]
   \label{eq:varphi}
\end{align}
where the second line follows by integration by parts.
This step can be iterated again as needs be, introducing an extra factor of $s^{-2}$ each time; such terms are negligible in the paraxial limit, and dominant for small $s$, as discussed below.

The longitudinal component ($B_z$) of Faraday's law $\nabla \times \bs{E} + \partial_t \bs{B} = 0$ gives
\begin{equation}
   \overline{\varphi} = \mp s^{-1} \left\{ \begin{smallmatrix} \partial_{\zeta} \\ \partial_{\zeta^*} \end{smallmatrix}\right\} \psi,
   \label{eq:curlEz}
\end{equation}
which gives a direct relation between the longitudinal magnetic field and $\psi$.

The other two helical components of Faraday's law, combined with (\ref{eq:curlEz}), give expressions for the remaining two fields
\begin{eqnarray}
   \overline{\psi} & = & \pm \left[ \mathrm{i} \psi - s^{-1} \left\{ \begin{smallmatrix} \partial_{\zeta^*} \\ \partial_{\zeta} \end{smallmatrix}\right\} \varphi + s^{-2} \partial_Z \psi \right],
   \label{eq:curlE-} \\
   \overline{\vartheta} & = & \pm s^{-1} \left\{ \begin{smallmatrix} \partial_{\zeta} \\ \partial_{\zeta^*} \end{smallmatrix}\right\} \varphi.
   \label{eq:curlE+}
\end{eqnarray}
The other Maxwell equations give no extra information.
Further manipulation can be used to show all of the fields satisfy the reduced Helmholtz equation
\begin{equation}
    0  =  (\partial_{\zeta}\partial_{\zeta^*} + \rmi \partial_Z  + \tfrac{1}{2}s^{-2}\partial^2_Z )\left\{\psi,\varphi,\overline{\psi},\overline{\vartheta},\overline{\varphi}\right\}, \label{eq:Hall}
\end{equation}
as indeed they must.

We first consider the paraxial regime $s \gg 1$.
(\ref{eq:Hall}) reduces to the paraxial equation $\partial_{\zeta}\partial_{\zeta^*}\bullet + \rmi \partial_Z \bullet = 0$.
Also, (\ref{eq:varphi}) gives that $\varphi \approx s^{-1} \partial_{\zeta}\psi$ or $s^{-1}  \partial_{\zeta^*}\psi$ and $\varphi = \pm \rmi \overline\varphi$.
The transverse components of the magnetic field are $\overline{\psi} = \pm \mathrm{i} \psi$, $\overline{\varphi} = \pm \mathrm{i} \varphi$ and $\overline{\vartheta} \sim O(s^{-2})$.
As expected, in the paraxial regime the longitudinal components of the beam ($z$) and the component of opposite handedness ($-$) are negligible compared to principal transverse components ($+$) and the electric and magnetic fields agree apart from a multiplicative factor that does not affect the vortex lines.

We now demonstrate that the symmetries of the torus knots' Milnor polynomials are transferred from $\psi$ to the other components of the beam. 
The trefoil knot field and most of the torus knot generalizations we consider have $\psi$ with the following form
\begin{equation} 
   \psi =  \zeta^n  + U(\zeta \zeta^{*},Z), 
   \label{eq:milnorform}
\end{equation}
where $U$ is some polynomial solution of (\ref{eq:Hall}) depending on $Z$ and $\zeta \zeta^*$.
$\psi$ satisfies (\ref{eq:Hall}), since $\zeta^n$ is a solution of the 2D Laplace equation; this part of the beam does not change on propagation.
For the case where $\psi$ multiplies $\widehat{\bs{e}}_+$, $\overline{\varphi}$ and paraxially $\varphi$ are proportional to $\partial_{\zeta^*}U = \zeta U'$ (with prime denoting derivative with respect to $\zeta\zeta^*$): an axisymmetric field times an axial vortex of strength 1.
This shows that the sign of OAM in $\psi$ has to be opposite the sign of the transverse circular polarization for the longitudinal component to have any degree of knot-like complexity in its amplitude structure.

On the other hand, in the case we mainly consider where $\psi$ multiplies $\widehat{\bs{e}}_-$, $\overline{\varphi}$ and paraxially $\varphi$ are proportional to $n \zeta^{n-1} + \zeta^* U' = \zeta^* [n \zeta^n + (\zeta \zeta^*)U']/(\zeta \zeta^*)$, which consists of an axial vortex of negative strength times a torus knot-like field with the same degree of rotational symmetry as $\psi$; this was indeed what was found for the trefoil knot case, and the properties generalize to the other torus knots quite directly.
We have not proved that the field $n \zeta^n + (\zeta \zeta^*)U'$ has a knot in its nodal structure, although it was in our main trefoil knot example, and in the other torus knots reported in Table 2 in the main text.

As $s$ varies into the nonparaxial regime, the longitudinal $B$-field retains the same relationship to $\psi$ as its derivative by (\ref{eq:curlEz}).
However, the relationship between the transverse and longitudinal $E$-field incurs more terms times $s^{-3}$, $s^{-5}$, \ldots. 
In the extreme nonparaxial limit $s \ll 1$, from (\ref{eq:varphi}) the dominant term in $\varphi$ is $\partial_{\zeta}\partial^N_Z \psi$ (where $N$ is the maximum integer which gives a nonzero expression): this gives $\zeta^*$ times an axisymmetric function, agreeing with Figure 3 (f) in the main text for the case of the trefoil.
An analogous argument can be made for $\overline{\psi}$ in the extreme nonparaxial limit, where $\overline{\psi} \approx -s^{-1}(\partial_{\zeta^*} + \partial_Z \psi)$, which for the knot function (\ref{eq:milnorform}) is axisymmetric, represented in Figure 3 (g) of the main text.

This discussion shows that the features discussed in the main text for the specific example of the trefoil knot are in fact more general, and apply to other torus knots whose $\psi$ function has the Milnor form (\ref{eq:milnorform}), specifically that :
\begin{itemize}
\item in the paraxial regime, all nonzero components have a nodal knot that follows the same curve in space; 
\item the longitudinal components have a negative strength axial vortex for transverse polarization $\widehat{\boldsymbol{e}}_-$, and a positive axial vortex for $\widehat{\boldsymbol{e}}_+$; 
\item the longitudinal magnetic component has a closer behaviour to the main transverse electric component that the longitudinal electric component; 
\item in the extreme nonparaxial limit, the transverse magnetic field becomes axisymmetric, and the longitudinal electric field becomes axisymmetric times a negative vortex;
\end{itemize}
where the last three cases hold for $\psi$ with positive OAM and polarization $\widehat{\boldsymbol{e}}_-$.

These facts all support the suggestion that in the crossover regime where $s \approx 1$, there will be a nontrivial nodal topology occupying a similar focal volume in all the components. 
The knot-like nodal lines in this bundle regime are shown in figure \ref{fig:ReconnectionMORE} (a) for $\varphi$, (g) for $\overline{\psi}$, and (j) for $\overline{\varphi}$ (which are trefoil knots) and $\overline{\vartheta}$ in (d) which, although possessing the right symmetries, is not knotted. 
Hence we anticipate a bundle structure for other kinds of torus knot; the knot type in their correspondent $E_z$ is reported in Table 2 of the main text.
The transition of the lines from the knot bundle topology to the topology of the extreme nonparaxial regime ($s \ll 1$) for $\varphi$, $\overline{\vartheta}$ $\overline{\psi}$, and $\overline{\varphi}$ is shown in figure \ref{fig:ReconnectionMORE}. 
The reconnections happen at different values of $s$ for the different components.

A slightly more general field, including the (3,2), (4,2) and (5,2) examples in Table 2, has the form $\zeta^m V(\zeta \zeta^*,Z) + U(\zeta \zeta^*,Z)$.
Fields with a knotted $\psi$ of this form do not quite have all the properties described above.
In particular, the axial component may be complicated and even knotted for $\psi$ if this form combined with $\widehat{\boldsymbol{e}}_+$.
In the examples in the table, the knot in the longitudinal component $\varphi$ is of a simpler torus knot type than the transverse field, reflecting the effects of $V(\zeta \zeta^*,Z)$ on the field.
We will not describe the properties of these fields further here, but it is straightforward to do so using similar analysis.

Finally, we note that our analysis here may be used to approach the problem described recently in [48] in the main text, which may be paraphrased in the present terminology as ``how can we determine $\psi$ such that $\varphi$ is a desired knotted field?'' 
Taking the desired $\varphi$ to be (\ref{eq:milnorform}), we can see immediately from (\ref{eq:divE}) that, for $\widehat{\boldsymbol{e}}_-$ polarization,
\begin{equation} 
   \psi = -\int^{\zeta} (\rmi s \varphi + s^{-1} \partial_Z\varphi)\rmd \zeta',
   \label{eq:phishape}
\end{equation} 
which can be calculated directly from our polynomial $\varphi$.
A cursory numerical investigation of this $\psi$ has an axial vortex and a nodal knot (as it must) for large $s$, but the knot structure breaks down for $s \approx 1$, leading to a system of concentric rings around the axial vortex.
We note that the approach in [48] assumed the transverse polarization was linear, hence the analysis there is somewhat more complicated, and the present approach based on helical states can readily be adapted to the linear transverse polarization case.

\end{document}